\documentclass[]{tMPH2e}

\usepackage{multirow}
\usepackage{array}
\usepackage{hhline}

\begin{document}


\articletype{ARTICLE}

\title{Fluorescence-lifetime-limited trapping of Rydberg helium atoms on a chip}

\author{V. Zhelyazkova$^{{a},{\dagger}}$\thanks{$^{\dagger}$Contributed equally. \vspace{6pt}}, M. \v Ze\v sko$^{{a},{\dagger}}$, H. Schmutz$^{a}$, J. A. Agner$^{a}$  and F. Merkt$^{a,{\ast}}$\thanks{$^\ast$Corresponding author. Email: merkt@phys.chem.ethz.ch 
\vspace{6pt}}\\ \vspace{6pt}  $^{a}${\em{Physical Chemistry Laboratory, ETH Z\"urich, 8093 Z\"urich, Switzerland}}}
\maketitle

\begin{abstract}
Metastable (1s)(2s) $^3{\rm S}_1$ helium atoms produced in a supersonic beam were excited to Rydberg-Stark states (with $n$ in the $27-30$ range) in a cryogenic environment and subsequently decelerated by, and trapped above, a surface-electrode decelerator. In the trapping experiments, the Rydberg atoms were brought to rest in 75~$\mu$s and over a distance of 33~mm and kept stationary for times $t_{\mathrm{trap}}$ in the $0-525$~$\mu$s range, before being re-accelerated for detection by pulsed field ionization. The use of a home-built valve producing short gas pulses with a duration of about 20~$\mu$s enabled the reduction of losses arising from collisions with atoms in the trailing part of the gas pulses. Cooling the decelerator to 4.7~K further suppressed losses by transitions induced by blackbody radiation and by collisions with atoms desorbing from the decelerator surface. The main contribution (60\%) to the atom loss during deceleration is attributed to the escape out of the decelerator moving traps of atoms having energies higher than the trap saddle point, spontaneous emission and collisions with atoms in the trailing part of the gas pulses causing each only about 20\% of the atom loss. At 4.7 K, the atom losses in the trapping phase of the experiments were found to be almost exclusively caused by spontaneous emission and the trap lifetimes were found to correspond to the natural lifetimes of the Rydberg-Stark states. Increasing the temperature to 100 K enhanced the trap losses by transitions stimulated by blackbody radiation.
\end{abstract}

\section{Introduction}
\label{introduction}

Atomic and molecular Rydberg states of high principal quantum number $n$ have large electric dipole moments, which in first approximation scale as $n^2$ for the outermost members of the Stark manifolds of states \cite{stebbings83a,gallagher94a}. Breeden and Metcalf \cite{breeden81a} and Wing \cite{wing80a} 
proposed to use these large electric dipole moments to accelerate and trap Rydberg atoms using inhomogeneous electric fields in the early 1980s. 

Softley and coworkers were the first to carry out experiments based on these early proposals and developed proof-of-principle devices with which they deflected and decelerated beams of Kr and Ar atoms \cite{townsend01a,vliegen04a} and of H$_2$ molecules \cite{procter03a,yamakita04a} excited to Rydberg states. Developing more complex electrode geometries soon enabled the trapping of translationally cold samples of hydrogen atoms \cite{vliegen07a,hogan08a} and molecules \cite{seiler09a,seiler11b}. 

In the past five years, these experiments have been extended to deflect, decelerate, split and trap beams of Rydberg atoms and molecules using chip-based devices \cite{hogan12b,allmendinger13a,allmendinger14a,palmer17a,lancuba16a}, to manipulate beams of cold positronium atoms \cite{alonso17a}, to study ion-molecule reactions at temperatures below 1~K \cite{allmendinger16a,allmendinger16b} and to investigate radiative processes and resonant collisions in dilute samples of cold Rydberg atoms \cite{seiler16a}. 

We present here experiments in which a cryogenic surface-electrode decelerator has been used in combination with a home-built valve generating short gas pulses \cite{motsch14a} to trap triplet Rydberg helium atoms under conditions where their decay by collisional processes and by transitions stimulated by blackbody radiation is suppressed. These experiments demonstrate a route towards measuring the natural lifetimes of Rydberg-Stark states of atoms and molecules.

 \section{Experimental procedure}
 \label{apparatus}
\begin{figure}[!h]
	\centering
	\includegraphics[width=0.9\linewidth]{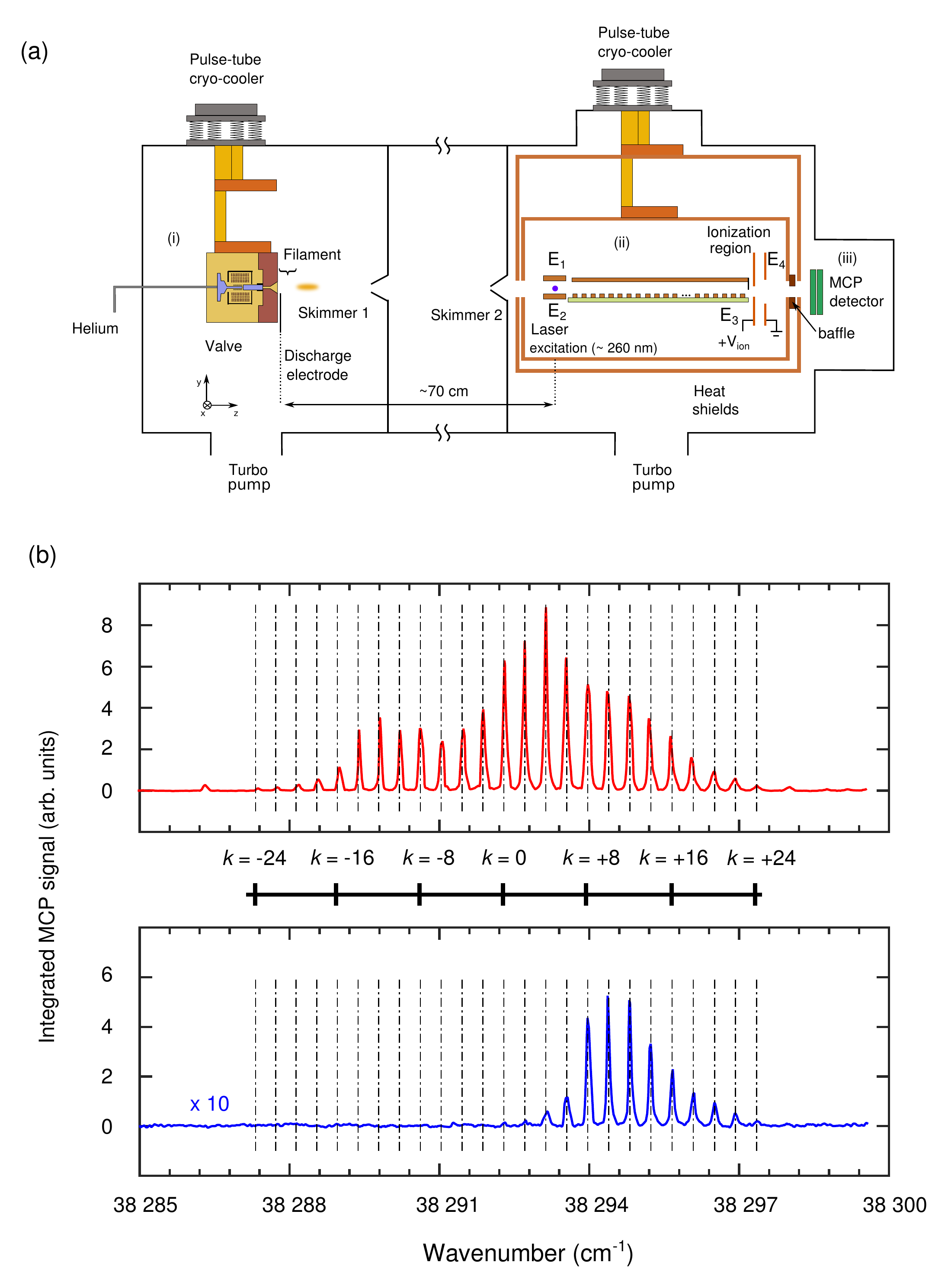}
	\caption{(a) Schematic view of the experimental set-up. (i) Source chamber with a home-built valve cooled to cryogenic temperatures used to produce the supersonic expansion. (ii) Excitation region, 44-electrode surface decelerator and ionization region cooled to cryogenic temperatures (in the $4.7-100$~K range) and enclosed in two heat shields in order to suppress blackbody radiation. (iii) Microchannel-plate (MCP) detector. (b) Excitation spectrum of the $n = 26,\,|m_\ell| = 1$ Rydberg-Stark manifold recorded from the (1s)(2s) $^3{\rm S}_1$ state in an excitation field of $F_{\mathrm{exc}}=125$\;V/cm, with the surface-electrode decelerator not active (red) and operated in deceleration mode to a final velocity of 400~m/s (blue). The calculated energy levels are shown as black dashed lines.}
	\label{exp-set-up}
\end{figure}

A schematic diagram of the experimental setup is presented in Fig.~\ref{exp-set-up}(a). A cold beam of metastable He atoms in the (1s)(2s)\;$^3$S$_1$ state (referred to as He* hereafter) is generated in the region labelled (i) by striking an electric discharge through a supersonic beam of pure He near the orifice of a pulsed valve. After collimation through two skimmers, the supersonic beam enters a second region, labelled (ii), located between a pair of parallel metallic plates [E$_1$ and E$_2$ in Fig.~\ref{exp-set-up}(a)], where the He* atoms are photoexcited in an electric field to selected Rydberg-Stark states. The Rydberg atoms in the beam then approach a surface-electrode Rydberg-Stark decelerator, where they can be guided, accelerated, decelerated, or trapped and then re-accelerated out of the trap for detection. After leaving the decelerator, the Rydberg atoms enter a region located between a second pair of parallel metallic plates (E$_3$ and E$_4$) where they are pulsed field ionized. The He$^+$ ions are extracted toward a microchannel-plate (MCP) detector [region (iii)] located at the end of a short field-free region beyond plate E$_4$.

The supersonic beam of ground-state He atoms is generated using a home-built pulsed valve operated at a repetition rate of 25 Hz and producing gas pulses with a duration of about 20~$\mu$s. The valve body is cooled with a two-stage pulse-tube cooler and counter-heated with two resistive elements to stabilise the temperature to within $\pm0.2\,\mathrm{K}$ in the $36-70$~K range, resulting in supersonic beams with mean forward longitudinal velocities in the $700-890$~m/s range. The electric discharge used to produce He* is ignited by applying a 40-$\mu$s-long potential pulse of +550\,V to a ring electrode located immediately after the nozzle orifice. This potential attracts electrons emitted by a hot tungsten filament, which seeds the discharge. The duration and delay of the potential pulse relative to the valve opening time are adjusted to optimize the density and stability of the He* atom beam.
The He* beam is skimmed twice before it enters the photoexcitation region, with the first (second) skimmer having a diameter of 19~mm (3~mm). The short nozzle-opening times and the strong cooling taking place in the supersonic expansion ensure that the gas pulses remain short during their propagation from the nozzle to the detection region. Measurements of the time-of-flight (TOF) distribution of the gas pulse indicate that the half width at half maximum of the gas pulse over the surface electrode decelerator is about 100~$\mu$s. This represents the first important advantage over our previous Rydberg-Stark deceleration and trapping experiments \cite{hogan12b,allmendinger13a}: The decelerated atoms are rapidly overtaken by all atoms in the trailing part of the pulse. Consequently, the trapped Rydberg atoms are not exposed to collisions with ground-state and metastable He atoms.

After the second skimmer, the He* atoms enter region (ii) [see Fig.~\ref{exp-set-up}(a)] where they are photoexcited to Rydberg-Stark states with principal quantum number $n$ around 30 using the frequency-tripled output of a commercial pulsed Nd:YAG-pumped dye laser (repetition rate 25~Hz, pulse duration $\sim$5~ns, fundamental wavelength $\sim$780~nm, bandwidth 0.05~cm$^{-1}$). The laser beam intersects the He* beam at right angles between two parallel copper plates [E$_1$ and E$_2$ in Fig.~\ref{exp-set-up}(a)] of dimensions 4~mm$\times 4$~mm separated by 4.8~mm in the $y$ direction, which are used to generate a static electric field $\vec{F}_{\mathrm{exc}}$ at excitation. The laser polarization can be set to be either parallel or perpendicular with respect to $\vec{F}_{\mathrm{exc}}$. In the former (latter) case, the He* atoms are photoexcited to Rydberg-Stark states with $m_\ell=0$ ($\pm 1$).
In the experiments presented here, $F_{\mathrm{exc}}$ was set such that the Rydberg-Stark states of interest were excited just below the Inglis-Teller (IT) limit, where the two manifolds of states with $n$ differing by one start to overlap. Near the IT limit, the separation between adjacent Rydberg-Stark states is large enough for individual Stark states to be resolved spectrally. 

The upper trace in Fig.~\ref{exp-set-up}(b) shows a typical excitation spectrum to $n=26,|m_\ell|=1$ He Rydberg-Stark states recorded in a field $F_{\mathrm{exc}}=125$~V/cm. The Stark states are labelled with the index $k$, which ranges from $-(n-|m_\ell|-1)$ to $(n-|m_\ell|-1)$ in steps of two \cite{gallagher94a}. The weak line located below the Stark manifold (at $\sim$38\;286.4~cm$^{-1}$) corresponds to an excitation to a state of dominant 25s, $m_\ell =0$ character that is observed because of a weak residual parallel component of the laser polarization. The Stark effect is essentially linear, with Stark shifts $E_{\rm Stark}^{n,k}$ given by 
\begin{equation}
E_{\rm Stark}^{n,k} = \frac{3}{2}a_0enkF_{\mathrm{exc}}. 
\end{equation}

To minimize losses by radiative decay, the Rydberg atoms enter the 4.5-cm-long surface-electrode decelerator immediately after photoexcitation. The decelerator consists of 44 electrodes [length 10~mm in the $x$ dimension, width 0.5~mm in the $z$ dimension, see Fig.~\ref{exp-set-up}(a)] printed on a circuit-board and separated by $d_z=1$~mm  in the beam-propagation ($z$) direction. Its operational principle is inspired from chip-based decelerators for polar molecules~\cite{meek08a,meek09a} and is described in Ref.~\cite{allmendinger13a}. Chirped sinusoidal electric potentials $V_i(t)$
\begin{equation}
V_i(t) = (-1)^i \, \frac{V_0}{2} \Bigg( 1 + \cos \Big[ \Big( \omega_0 +  \omega^\prime(t-t_0) \Big) (t-t_0) + \phi_i \Big] \Bigg)
\label{eq1}
\end{equation}
are applied to each electrode to create a regular array of tubular electric quadrupole traps that move above the surface of the decelerator with an initial velocity $v_{z,0}= 3 d_z\omega_0/(2\pi)$ and a constant acceleration $a_z = 3 d_z\omega^\prime/(2\pi)$. 
In Eq.~(\ref{eq1}), $i = 1,\ldots,44$ is the electrode index, $\phi_i = (2-i) \cdot \frac{2\pi}{3}$, and $t_0$ is the time at which the potentials are turned on. The amplitude $V_0$ is chosen in the range $40-80$~V, depending on the selected $n$ and $k$ values of the excited Rydberg-Stark states, to ensure a tight-enough trap while avoiding electric fields above the IT limit which cause transitions between Stark states of opposite $k$ values~\cite{vliegen04a}. The size of the electrodes is designed so that all atoms excited to Rydberg-Stark states can be loaded into a single trap. The switch-on time of the potentials $t_0$ is chosen to maximize the number of Rydberg atoms loaded into the first trap of the array. 

The potentials $V_i(t)$ are generated using an arbitrary waveform generator and subsequently amplified before being applied to the electrodes. Because of the periodicity of the electrode structure, only six potential functions are required, each connected to one of the six subsets labelled by $j=1-6$, where $j=(i\mod6)+1$. The electric-field minima generated by the oscillating electric potentials are situated approximately 0.75~mm above the chip surface and separated by 3~mm in the beam-propagation direction. 

The forces acting on a Rydberg atom and resulting from the inhomogeneous electric-field distribution generated by the decelerator electrodes are given to first order by
\begin{equation}
-\nabla E_{\rm Stark}^{n,k} = -\frac{3}{2} nkea_0 \nabla F,
\label{eq2}
\end{equation}
where $\nabla F$ is the gradient of the electric field at the position of the atom. 
The choice of $n$ and $k$ values of the Stark states strongly affects the photoexcitation yield and the efficiency of the deceleration process. They were chosen to ensure (i) sufficiently large photoexcitation cross sections from the metastable (1s)(2s) $^3$S$_1$ state, which scale as $n^{-4}$ for Rydberg-Stark states, (ii) sufficiently long lifetimes, which scale as $n^{4}$ and are typically longer than 50~$\mu$s at $n\geq 25$ and for blackbody temperatures $T_{\rm BB}$ below 100~K (see Sec.~\ref{trapRes}), and (iii) sufficiently large confining forces, which scale as $nk$. 

The $k$ dependence of the photoexcitation probabilities and lifetimes also needs to be considered. The bottom trace in Fig.~\ref{exp-set-up}(b) illustrates these considerations. It displays the Rydberg-atom signal measured after deceleration of $n=26,|m_\ell|= 1$ He Rydberg Stark states from an initial velocity of about 700~m/s to a final velocity of 400~m/s with $V_0=80$~V. No signal is observed for high-field-seeking Stark states (states with $k\leq 0$) and the maximal signal is observed for $k=10$. States on the edge of the manifold have lower photoexcitation probabilities which rules out the use of Stark states with maximal $k$ value. The best choice of $k$ thus represents the optimal compromise between photoexcitation and deceleration efficiency.

After the decelerator, the Rydberg atoms enter a region where they are detected by pulsed field ionization (PFI). This region is confined by two parallel copper plates [E$_3$ and E$_4$ in Fig.~\ref{exp-set-up}(a)] separated by 15~mm in the $z$ direction. PFI is achieved by applying a pulsed potential ($V_{\mathrm{ion}}=+1.5$~kV, rise time 50~ns, duration $\sim$500~ns) to electrode E$_3$ and keeping electrode E$_4$ grounded. The produced He$^+$ ions are accelerated toward an MCP detector located $\sim$50~mm beyond the PFI region. The TOF distribution of the Rydberg atoms from the photoexcitation spot to the PFI region is determined by monitoring the PFI signal as a function of the delay between the photoexcitation laser pulse and the application of the field-ionization pulse. From this TOF distribution, the velocity distribution of the Rydberg atoms at the end of the decelerator can easily be reconstructed, as explained in Ref.~\cite{allmendinger13a}.
Because the experimental geometry is precisely known, the deceleration process for a chosen set of initial velocity $v_{z,0}$ and acceleration $a_z$ can be quantitatively predicted by Monte-Carlo particle-trajectory simulations. The predictions can also be validated by comparing calculated and measured TOF distributions, as demonstrated in Ref.~\cite{allmendinger13a} (see also Fig.~\ref{guiding-fig} below). 

In the trapping experiments presented in this article, the Rydberg atoms are first decelerated to zero velocity at approximately 3/4 of the total length of the decelerator. After this deceleration phase, the quadrupole trap and the Rydberg atoms contained in it are kept stationary above the decelerator surface for a predefined amount of time $t_{\mathrm{trap}}$, after which they are re-accelerated to a velocity of 400~m/s at the end of the decelerator. 
The Rydberg atoms then fly into the region where they are detected by PFI. Because the deceleration and re-acceleration phases are identical in all trapping experiments, monitoring the detected Rydberg-atom signal as a function of $t_{\mathrm{trap}}$ provides information on the Rydberg-atom decay during the stationary phase.  

To suppress trap losses caused by blackbody-radiation-induced transitions and photoionization, the decelerator, excitation electrodes and field-ionization plates can be cooled to temperatures in the $4.7-100$~K range by a two-stage pulse-tube cooler. Moreover, two heat shields and a baffle are used to reduce the contribution from the thermal radiation of the room-temperature laboratory environment to below 1\%. This represents the second advantage over our previous experiments \cite{allmendinger13a} and enables us to disentangle the different collisional and radiative trap-loss processes.

\section{Results}
\subsection{Guiding and deceleration}
\label{guideDecelRes}
The surface-electrode decelerator can be operated in three modes: (i) a guiding mode, in which the electric quadrupole traps move at a constant velocity $v_{\mathrm{guide}}$, (ii) a deceleration or acceleration mode, in which the quadrupole traps are uniformly decelerated or accelerated to a desired final velocity $v_f$, and (iii) a trapping mode, in which the atoms are decelerated to zero velocity, kept stationary for a specified amount of time $t_{\mathrm{trap}}$ and subsequently re-accelerated for detection. In this subsection, we present the results of experiments performed to characterize the Rydberg-atom losses and the final velocity distributions resulting from the guiding and deceleration processes. These experiments differ from those presented in Ref.~\cite{allmendinger13a} through (i) the different gas-pulse characteristics that result from the much shorter valve-opening times, i.e., 20~$\mu$s instead of 200~$\mu$s in Ref.~\cite{allmendinger13a}, (ii) the longer flight distance between the nozzle and laser-excitation spot, i.e., 70~cm instead of about 30~cm, (iii) the lower valve temperatures used here resulting in lower initial beam velocities, i.e. 700-800~m/s instead of 1200~m/s, and (iv) the possibility to almost entirely suppress blackbody radiation. 

\begin{figure}[!h]
	\centering
	\includegraphics[width=1.0\textwidth]{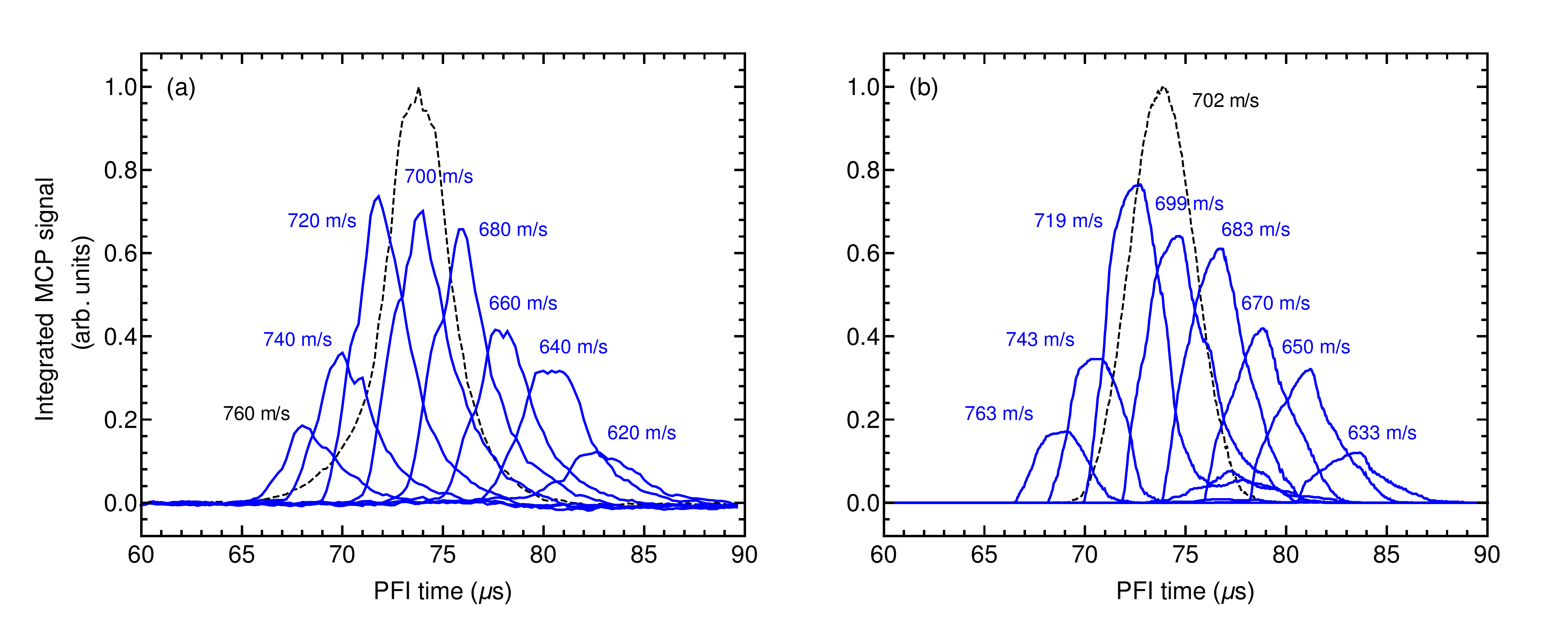}
	\caption{(a) Recorded TOF profiles of the helium atoms exited to the $|n,k,m\rangle=|30,24,0\rangle$ state as a function of the time between excitation and the application of the PFI pulse, with no waveforms applied to the decelerator (dashed black line) and with the chip operated in guiding mode with set velocities in the $v_{\mathrm{guide}}=620-760$~m/s range, as indicated above each trace. (b) Corresponding simulated TOF profiles with a particle-trajectory simulation. The  labels above each trace correspond to the mean velocities of the particles extracted from the simulation before (black) and after (blue) the application of the guiding waveforms.}
	\label{guiding-fig}
\end{figure}
The experimentally measured TOF profiles obtained after guiding He Rydberg atoms in the $|n,k,m_\ell \rangle=|30,24,0\rangle$ state with the surface-electrode decelerator cooled to 4.7~K [panel (a)] is compared in Fig.~\ref{guiding-fig} with the results of numerical particle-trajectory simulations [panel (b)]. The experimental profiles were obtained by monitoring the PFI signal as a function of the time delay between laser-excitation and the field-ionisation pulse. 
The selected guiding velocities $v_{\mathrm{guide}}$ are indicated above each solid blue trace and range from 760 to 620~m/s. For comparison, the TOF profile recorded without applying any potential to the decelerator electrodes is depicted as dashed black line. Its peak intensity is located at 74.9~$\mu$s  and corresponds to a Rydberg-atom beam with a mean velocity of $v_0=702$~m/s and a full-width-at-half-maximum (FWHM) velocity spread in the beam propagation direction of 47~m/s, as determined from the simulation. This distribution is significantly narrower than the analogous distribution reported in Fig.~4 of Ref.~\cite{allmendinger13a}, which is a consequence of the ten-times shorter gas-pulse duration: The different velocity classes of a short initial gas pulse disperse more between the nozzle orifice and the laser-excitation spot, and consequently the laser pulse excites atoms with a narrower velocity distribution. 

As expected, the selection of guiding velocities that are higher (lower) than $v_0$ shifts the intensity maxima of each measured TOF profile to earlier (later) times of flight. For instance, Rydberg atoms guided at 760~m/s (620~m/s) have a central arrival time of 67.9~$\mu$s (82.4~$\mu$s). For any value of $v_{\mathrm{guide}}$ different from 700~m/s, the signal intensity is larger than that of the original beam at the corresponding time of flight, as can be easily seen by comparing the dashed black and the full blue lines in Fig.~\ref{guiding-fig}(a). This intensity enhancement can be explained by the fact that, when operated in guiding mode, the surface-electrode decelerator also acts as an accelerator or decelerator for atoms travelling at velocities close to $v_{\mathrm{guide}}$. This effect is quantitatively accurately described by the Monte-Carlo particle-trajectory simulations presented in Fig.~\ref{guiding-fig}(b), from which one can extract the actual velocity distributions at the end of the surface-electrode decelerator.

The mean velocities of the beam after the decelerator agree within the statistical uncertainties (about $\pm 10$~m/s) with the selected values of $v_{\mathrm{guide}}$. However, guiding results in heating of the Rydberg-atom cloud. Whereas the velocity distribution prior to the surface-electrode decelerator is well described by a longitudinal temperature of 65~mK and an even colder transverse temperature of less than 1~mK, the velocity distributions after guiding above the surface-electrode decelerator indicate equal widths in the longitudinal and the transverse directions and correspond to a translational temperature of about 1.5~K. The values of $V_0$ used to obtain the data presented in Fig.~\ref{guiding-fig} correspond to a trap depth of 2.4~K with respect to the saddle point of the moving traps for the $|n,k,m\rangle=|30,24,0\rangle$ Rydberg-Stark state. In the present experiment, $t_0$ is optimized experimentally so that the Rydberg atoms are loaded near the centre of the trap. The simulations indicate that the final velocity distributions after guiding are determined by: (i) the initial size of the Rydberg-atom cloud, which is itself given by the intersection volume of the laser and gas beams, and (ii) the coupling between longitudinal and transverse motion in the moving traps. Colder samples can be obtained by reducing the value of $V_0$ and better collimating the laser and gas beams, however, at the expense of the total number of trapped atoms, as discussed in more detail in Ref.~\cite{allmendinger14a}. 

Deceleration is typically accompanied by losses of Rydberg atoms through (i) spontaneous (ia) and stimulated (ib) radiative processes, (ii) collisions, and (iii) through the escape out of the moving traps of atoms having energies higher than the trap saddle point of the effective trap potential associated with the moving reference frame corresponding to the electric-field minimum~\cite{hogan13a}. The collisional losses themselves can have different origins: (iia) collisions with background-gas particles in the chamber, (iib) collisions with atoms in the trailing edge of the gas pulse \cite{seiler11a}, (iic)  Rydberg-Rydberg collisions in the moving traps, and (iid) collisions with particles desorbing from the surface electrodes of the decelerator, as suggested in Ref.~\cite{lancuba16a}.

The cryogenic environment used in the present experiments enables the suppression of atom losses through stimulated radiative transitions \cite{seiler16a} (ib), and through collisions with background-gas particles (iia). Desorption from the surface electrodes (iid) is also suppressed by cooling the decelerator to 4.7~K. The use of a short gas pulse, with half width at half maximum of less than 100~$\mu$s above the decelerator, further ensures a strong reduction of collisions with atoms in the trailing part of the gas pulses after about 100~$\mu$s (iib). Finally, the experiments were performed at Rydberg-atom densities below $10^4$~atoms/cm$^3$, which suppresses losses by Rydberg-Rydberg collisions \cite{seiler16a} (iic).

The black dots with statistical error bars ($1\sigma$) in Fig.~\ref{trap-fig}(a) display the relative total integrated Rydberg-atom signal monitored for a range of PFI times at the end of the decelerator following deceleration from an initial velocity of 890~m/s to final velocities $v_{\rm f}$ in the $880-150$\;m/s range (indicated above each data point). The atoms were initially excited to the  $\left|27,18,0\right\rangle$ state and the decelerator and its environment were cooled to 4.7~K. The initial mean longitudinal velocity of the unperturbed Rydberg-atom beam was estimated from the particle-trajectory simulations to be 890\,m/s. The accelerations used in these experiments ranged from $-2.16\times10^5$~m/s$^{2}$ for $v_{\rm f}=880$~m/s to $-9.39\times10^6$~m/s$^{2}$ for $v_{\rm f}=150$~m/s, and the corresponding deceleration sequence durations ranged from 48~$\mu$s to 80~$\mu$s, which are less than the half-width (about 100~$\mu$s) of the gas pulse duration above the decelerator. The dashed black line in Fig.~\ref{trap-fig}(a) is an exponential fit to the data points with a characteristic decay constant of 21.1~$\mu$s, and the solid green line represents the exponential decay associated with the radiative lifetime of the $|27,18,0\rangle$ state, calculated to be 104~$\mu$s at 4.7~K, as explained in Section~\ref{discussion}.

From Fig.~\ref{trap-fig}(a) it is apparent that the total detected signal rapidly decreases with increasing deceleration. The experimental conditions chosen for these experiments rule out most decay mechanisms and only leave contributions from mechanisms (ia), (iib) and (iii). The analysis of this experiment with numerical particle-trajectory simulations indicates that (iii) is the dominant loss mechanism and accounts for 60\% of the total loss observed in the deceleration experiment to $v_{\rm f}=150$~m/s, the losses from spontaneous emission and collisions with particles in the trailing part of the pulse being each about 20\%. The losses from collisions with atoms in the trailing part of the gas pulse are strongly reduced compared to the losses observed when using longer nozzle-opening times \cite{allmendinger13a} because of the reduced total amount of gas emitted by the nozzle and the associated gas-density reduction resulting from the velocity dispersion. 

\subsection{Trapping}
\label{trapRes}

The measurements of the decay of trapped Rydberg atoms were performed with He Rydberg atoms initially travelling at a velocity of $v_{\rm i}=880$~m/s. The experiment consisted of (i) a deceleration phase to zero final velocity, with a deceleration of $a_1=-1.2\times10^7$~m/s$^{2}$ and a deceleration sequence time of about 73~$\mu$s, (ii) a stationary phase, during which the traps were held fixed in space for a selected time $t_{\mathrm{trap}}$ in the $0-525$~$\mu$s range, and (iii) a re-acceleration phase to a final velocity of 400~m/s (acceleration $a_2=1.1\times10^7$~m/s$^{2}$) for detection. The total duration of the sequence applied to the chip electrodes in the case of $t_{\mathrm{trap}}=0$~$\mu$s was 112~$\mu$s.  
\begin{figure}[!h]
	\centering
	\includegraphics[width=1.0\textwidth]{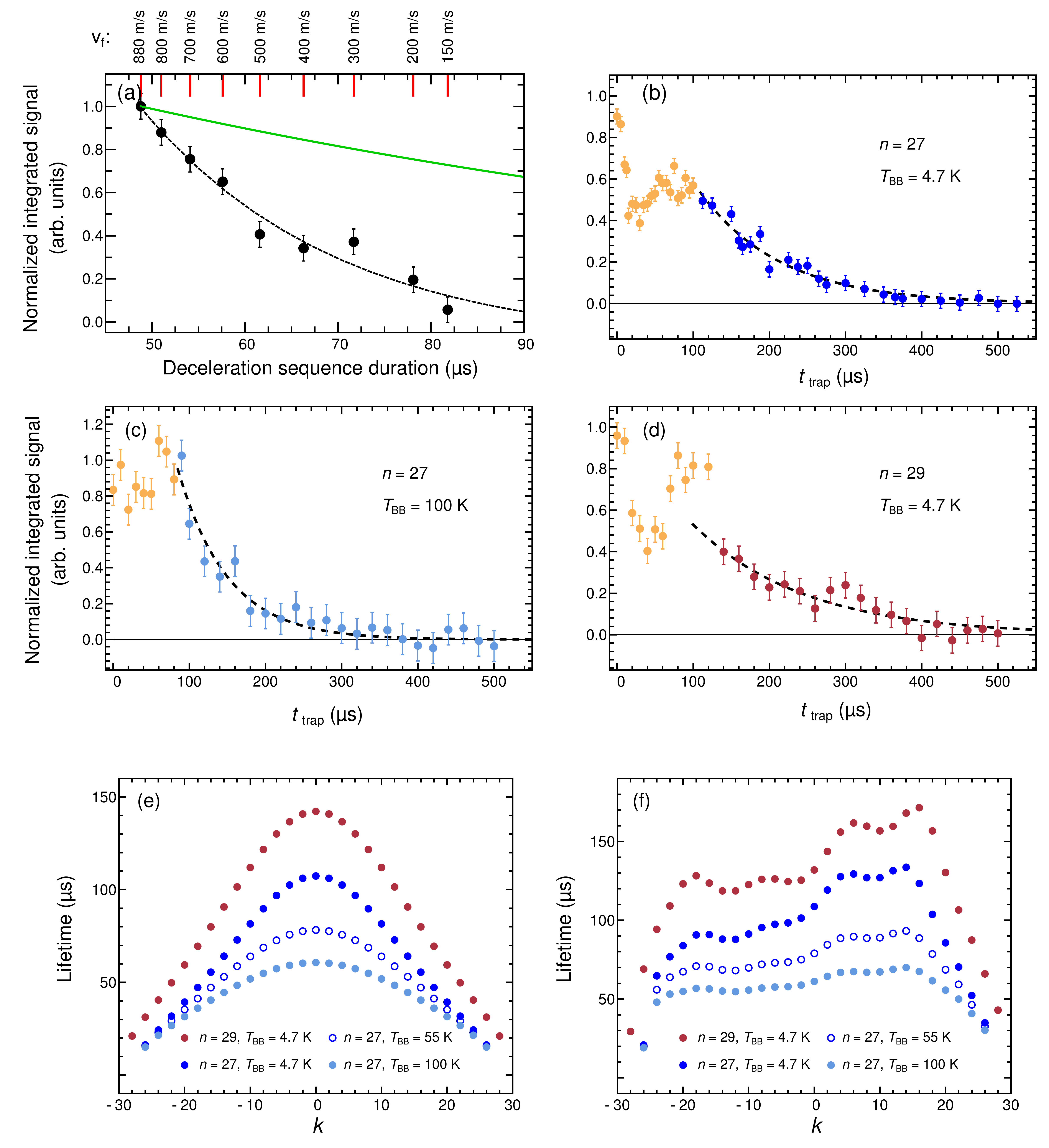}
	\caption{(a) Integrated TOF profiles for the helium atoms excited to the $|27,18,0\rangle$ state decelerated from an initial velocity of 890~m/s to final velocities $v_{\rm f}$ in the $880-150$~m/s range, as indicated above each point, as a function of the deceleration-sequence duration. The chip environment is cooled to $T_{\mathrm{BB}}=4.7$~K. Shown in the dashed black curve is the exponential decay fit to the data points (time constant of 21.1~$\mu$s), and in the solid green line the exponential decay associated with the fluorescence lifetime of the state (calculated to be 104~$\mu$s). (b), (c) and (d) The integrated TOF profiles of the helium atoms excited to the $|27,18,0\rangle$ state with the surface--electrode decelerator cooled to $T_{\mathrm{BB}}=4.7$ and 100~K [(b) and (c), respectively] and excited to the $|29,18,0\rangle$ state with $T_{\mathrm{BB}}=4.7$~K [(d)], and trapped for $t_{\mathrm{trap}}=0-525$~$\mu$s. The experimental data points are averages over 25 cylcles. The integrated signal decay after the initial oscillations at $t_{\mathrm{trap}}\lesssim90$~$\mu$s is fitted to an exponential decay (dashed black lines). The data points used for each fit are displayed in dark blue (b), light blue (c) and red (d). The calculated lifetimes of the $n=27$ and $n=29$, $m_\ell=0$ Rydberg-Stark manifolds in hydrogen (e) and the triplet states of helium (f), at temperatures as indicated.} \label{trap-fig}
\end{figure}

Panels (b) and (d) of Fig.~\ref{trap-fig} present the relative Rydberg-atom signal monitored by PFI as a function of $t_{\mathrm{trap}}$ after preparation of the atoms in the $|27,18,0\rangle$ and $|29,18,0\rangle$ Rydberg-Stark states, respectively. In both cases, the surface-electrode decelerator and its environment were kept at $T_{\mathrm{BB}}= 4.7$~K. For comparison, panel (c) shows the results obtained after preparation of the atoms in the $|27,18,0\rangle$ state but at $T_{\mathrm{BB}}=100$~K. In all three data sets, the signal reveals an oscillatory behaviour at early times up to about 100~$\mu$s. On the basis of the particle-trajectory simulations, we attribute this behaviour to a rapid equilibration of the translational motion resulting from the atoms filling the phase-space inside the trap. This initial phase is followed by a second phase in which the signal decreases exponentially, as indicated by the dashed black lines, which correspond to fitted exponential decay functions. The trapping times (approximately 100~$\mu$s) after which the data points included in the fits were chosen so as to minimize the statistical uncertainty, and correspond to the region where no effects from the initial phase-space equilibration dynamics are detectable anymore. They are also longer than the half width of the gas pulse, which enables us to rule out significant contributions to the decay from collisions with atoms in the late part of the gas pulse. The data points included in each fit are depicted in dark blue in Fig.~\ref{trap-fig}(b), light blue in Fig.~\ref{trap-fig}(c) and red in Fig.~\ref{trap-fig}(d).

The exponential (1/e) decay times $\tau_{\mathrm{meas}}$ obtained from the fits are listed in Table~\ref{table_LT}. 
The statistical uncertainties represent one standard deviation of the fitted values and the systematic uncertainties reflect the influence of the background level at late trapping times. \\   
\begin{table}	
	\centering
	
	\begin{tabular}{|c|c|c|}
		
		\hline
		
		\rule{0pt}{18pt}& $T_{\mathrm{BB}}$ = 4.7~K &   $T_{\mathrm{BB}}$ = 100~K \\
		&  &  \\
		\hline
		&  &  \\	
		\rule{0pt}{-1pt}	$|27,18,0\rangle$ &  &  \\ 
		$\tau_{\mathrm{meas}}$ &      ($106\pm8_{\mathrm{syst}}\pm8_{\mathrm{stat}}$)\;$\mu$s  & ($66\pm7_{\mathrm{syst}}\pm13_{\mathrm{stat}}$)\;$\mu$s \\ 
		$\tau_\mathrm{{calc}}$ &      103.7~$\mu$s   & 61.6~$\mu$s \\
		\hline
		
		\rule{0pt}{12pt} $|29,18,0\rangle$ & &   \\ 
		$\tau_{\mathrm{meas}}$ &      ($145\pm8_{\mathrm{syst}}\pm35_{\mathrm{stat}}$)\;$\mu$s   & -- \\ 
		$\tau_\mathrm{{calc}}$ &      156.7~$\mu$s     &  \\
		\hline
	\end{tabular} 
	\caption{Comparison of calculated lifetimes ($\tau_{\mathrm{calc}}$) and observed trap-decay times ($\tau_{\mathrm{meas}}$) of the $\left|27,18,0\right\rangle$ and $\left|29,18,0\right\rangle$ triplet Rydberg-Stark states of He at temperatures of $T_{\rm BB}$ of 4.7~K and  100~K.}\label{table_LT}
\end{table}  
These observed trap-decay times are much longer than those observed at $n=30$ at room temperature in Ref.~\cite{allmendinger13a} and decrease with increasing environment temperature. Transitions induced by blackbody radiation thus limit the trapping times at elevated temperatures, which confirms that trap-loss processes by collisions are strongly suppressed in the present experiments. The lifetimes closely correspond to the radiative lifetimes, as discussed in the next section.

Unfortunately, the experimental uncertainties are too large for a clear difference in lifetimes between the $\left|27,18,0\right\rangle$ and $\left|29,18,0\right\rangle$ triplet Rydberg-Stark states of He to be observable. The statistical uncertainties are limited by the small number of detected PFI-events and the systematic uncertainty caused by a weak background signal, as mentioned above. The largest PFI-signal was measured for the case in which the atoms were excited to the $|27,18,0\rangle$ state and the chip was cooled to 4.7~K [Fig.~\ref{trap-fig}(b)]. In the data set presented in Fig.~\ref{trap-fig}(b), the number of PFI events detected per experimental cycle at $t_{\mathrm{trap}}=0$ and 525~$\mu$s is about 30 and 2, respectively. Because of the loss of atoms during the re-acceleration for detection, this number represents a lower limit of the number of trapped atoms. In the other two data sets [Fig.~\ref{trap-fig}(c) and (d)], the PFI signal was further reduced by the enhanced decay rate at 100~K in the case of the second measurement carried out with atoms in the $\left|27,18,0\right\rangle$ Rydberg Stark state and by the lower excitation probability in the case of measurement involving the $|29,18,0\rangle$ Rydberg-Stark state, for which no measurement of sufficient quality could be performed at 100~K.

\section{Discussion and conclusions}
\label{discussion}
To calculate the radiative lifetimes of the triplet Rydberg-Stark states of He, we followed the procedure presented in detail in Ref.~\cite{seiler16a} and which led to a complete analysis of radiative and collisional decay process of Rydberg-Stark states of atomic hydrogen. The procedure consists of determining: (i) the Einstein $A$ coefficients for spontaneous emission from the selected Rydberg-Stark state $|nkm\rangle$ to all lower-lying states in the electric-dipole approximation and summing them to obtain an overall spontaneous decay rate $k_{\rm sp}$; (ii) the Einstein $B$ coefficients describing the rates of stimulated transitions to neighbouring states for a given blackbody-radiation temperature and summing them in order to obtain an overall stimulated decay rate $k_{\rm st}$; and (iii) taking the inverse of the sum $k_{\rm tot}=k_{\rm sp}+k_{\rm st}$ to obtain the lifetime $\tau_{nkm}=1/k_{\mathrm{tot}}$. The main differences compared to the calculations carried out for H in Ref.~\cite{seiler16a} are that the radial parts of the electric dipole transition moments were evaluated numerically using the Numerov method as described in Ref.~\cite{zimmerman79a} with the quantum defects reported in Ref.~\cite{drake91a}, and that we ignored blackbody-radiation-induced ionization transitions because of their negligible contribution at the $n$ values and temperatures of interest. 

The calculated radiative lifetimes of the $n=27$ and $n=29$, $m_\ell=0$ Rydberg-Stark states of hydrogen and triplet helium at temperatures of 4.7, 55 and 100~K are compared in Fig.~\ref{trap-fig}(e) and (f). In the case of H, the lifetimes exhibit a strong dependence on $k$, with states in the middle of the manifold being significantly longer-lived than the ones on the edges. For example, in the $n=27$ manifold at a temperature of 4.7~K the $k=0$ state of H has a radiative lifetime of 107.3~$\mu$s, whereas the outermost states of the Stark manifold with $|k|=26$ have a lifetime of 16.2~$\mu$s. This effect can be explained by the fact that the states located at the edges of the manifold have the largest $\ell=1$ character and thus decay most efficiently to the ground (1s) $^2$S$_{1/2}$ state. In the triplet states of helium, the $k$ dependence of the radiative lifetimes has slightly different characteristics resulting from the non-zero quantum defects of the low-$\ell$ states, which introduces an asymmetry around $k=0$. Stark states with negative Stark shifts (negative $k$ values) gain more low-$\ell$ character than Stark states with positive energy shifts (positive $k$ values) and are thus shorter-lived. Because the degree of Stark mixing with low-$\ell$ states increases with increasing field, the Stark state lifetimes are field dependent. The lifetimes presented in Fig.~\ref{trap-fig}(f) were calculated for a dc field of 40~V/cm corresponding to the average field in the trap. The calculations show that the natural lifetimes of the $n=27$ and 29, $k=18$ states do not depend strongly on the electric field strength. In addition, because the spontaneous decay rates are dominated by the contribution to the lowest-lying optically accessible level and the transitions from the triplet (1s)($nk$) $^3$S$_1$ Rydberg-Stark states to the ground (1s)$^2$ $^1$S$_0$ singlet state are forbidden, the lifetimes of the triplet Rydberg-Stark states in helium are longer than those of hydrogen. The differences of radiative lifetimes between H and He decrease with increasing temperature, because the contributions of blackbody-radiation-induced transitions, which are similar in both cases, become dominant.

The calculated radiative lifetimes ($\tau_{\mathrm{calc}}=1/k_{\mathrm{tot}}$) of the $|27,18,0\rangle$ and $|29,18,0\rangle$ Rydberg-Stark states of triplet He at temperatures of 4.7 and 100~K are presented in Table~\ref{table_LT}, where they are compared to the trap $1/\mathrm{e}$ decay times determined from the experimental data, as described in Section~\ref{trapRes}. In all cases, the calculated lifetimes are well within one standard deviation of the values of $\tau_{\mathrm{meas}}$. 

The number of trapped atoms implies that the initial density in the trap is on the order of $5\times10^3$~particles~cm$^{-3}$. In this density regime, $k$- and $m_\ell$-changing collisions mediated by resonant dipole-dipole interactions are negligible, as explained in Ref.~\cite{seiler16a}. These transitions typically populate high-$m_\ell$ states and strongly enhance trap-decay times. In the experiments presented here, the observed trap lifetimes are limited by radiative processes. In the measurements carried out at 4.7~K, where transitions stimulated by blackbody radiation are negligible, $\tau_{\mathrm{meas}}$ closely corresponds to the fluorescence lifetimes. 

The use of short gas pulses in combination with a cryogenic environment thus enables one to suppress contributions to the decay of trapped Rydberg atoms from collisional and blackbody-radiation-induced transitions to the extent that the measured trap-decay curves represent the natural lifetimes of the Rydberg-Stark states investigated. In studies of molecular Rydberg-Stark states, similar measurements would provide access to the determination of predissociation and autoionization lifetimes, and this aspect represents an important objective for future work.

\section*{Acknowledgments}
We dedicate this article to Timothy P. Softley and thank him for his inspiring role over many years and for his interest in the research presented in this article. This work was supported financially by the Swiss National Science Foundation (Grant No. 200020-172620), the Swiss national competence center for research (NCCR QSIT), and the European Research Council (ERC) under the European Union's Horizon 2020 research and innovation programme (Advanced Grant No. 743121).

\begin{thebibliography}{28}
	\providecommand{\url}[1]{\texttt{#1}}
	\providecommand{\urlprefix}{URL }
	
	\bibitem{stebbings83a}
	R.F. Stebbings and F.B. Dunning, editors, \emph{{Rydberg States of Atoms and
			Molecules}}   (Cambridge University Press, Cambridge, 1983).
	
	\bibitem{gallagher94a}
	T.F. Gallagher, \emph{{Rydberg Atoms}}   (Cambridge University Press,
	Cambridge, 1994).
	
	\bibitem{breeden81a}
	T. Breeden and H. Metcalf,  Phys. Rev. Lett.  \textbf{47}, 1726--1729
	(1981).
	
	\bibitem{wing80a}
	W.H. Wing,  Phys. Rev. Lett.  \textbf{45}, 631--634 (1980).
	
	\bibitem{townsend01a}
	D. Townsend, A.L. Goodgame, S.R. Procter, S.R. Mackenzie and T.P. Softley,  J.
	Phys. B: At. Mol. Opt. Phys.  \textbf{34}, 439--450 (2001).
	

	\bibitem{vliegen04a}
	E. Vliegen, H.J. W{\"o}rner, T.P. Softley and F. Merkt,  Phys. Rev. Lett.
	\textbf{92}, 033005 (2004).
	
	\bibitem{procter03a}
	S.R. Procter, Y. Yamakita, F. Merkt and T.P. Softley,  Chem. Phys. Lett.
	\textbf{374}, 667--675 (2003).
	
	\bibitem{yamakita04a}
	Y. Yamakita, S.R. Procter, A.L. Goodgame, T.P. Softley and F. Merkt,  J. Chem.
	Phys.  \textbf{121}, 1419--1431 (2004).
	
	\bibitem{vliegen07a}
	E. Vliegen, S.D. Hogan, H. Schmutz and F. Merkt,  Phys. Rev. A  \textbf{76}, 023405 (2007).
	
	\bibitem{hogan08a}
	S.D. Hogan and F. Merkt,  Phys. Rev. Lett.  \textbf{100}, 043001 (2008).
	
	\bibitem{seiler09a}
	S.D. Hogan, {\mbox{Ch}}. Seiler,  and F. Merkt,  Phys. Rev. Lett.  \textbf{103}, 123001 (2009).
	
	\bibitem{seiler11b}
	{\mbox{Ch}}. Seiler, S.D. Hogan and F. Merkt,  Phys. Chem. Chem. Phys.
	\textbf{13}, 19000--19012 (2011).
	
	\bibitem{hogan12b}
	S.D. Hogan, P. Allmendinger, H. Sassmannshausen, H. Schmutz and F. Merkt,
	Phys. Rev. Lett.  \textbf{108}, 063008 (2012).
	
	\bibitem{allmendinger13a}
	P. Allmendinger, J.A. Agner, H. Schmutz and F. Merkt,  Phys. Rev. A
	\textbf{88}, 043433 (2013).
	
	\bibitem{allmendinger14a}
	P. Allmendinger, J. Deiglmayr, J.A. Agner, H. Schmutz and F. Merkt,  Phys. Rev.
	A  \textbf{90}, 043403 (2014).
	
	\bibitem{palmer17a}
	J. Palmer and S.D. Hogan,  Phys. Rev. A  \textbf{95}, 053413 (2017).
	
	\bibitem{lancuba16a}
	P. Lancuba and S.D. Hogan,  {J. Phys. B: At. Mol. Opt. Phys.}  \textbf{49},
	074006 (2016).
	
	\bibitem{alonso17a}
	A.M. Alonso, B.S. Cooper, A. Deller, L. Gurung, S.D. Hogan and D.B. Cassidy,
	Phys. Rev. A  \textbf{95}, 053409 (2017).
	
	\bibitem{allmendinger16a}
	P. Allmendinger, J. Deiglmayr, O. Schullian, K. H{\"{o}}veler, J.A. Agner, H.
	Schmutz and F. Merkt,  ChemPhysChem  \textbf{17}, 3596--3608 (2016).
	
	\bibitem{allmendinger16b}
	P. Allmendinger, J. Deiglmayr, K. H{\"{o}}veler, O. Schullian and F. Merkt,  J.
	Chem. Phys.  \textbf{145}, 244316 (2016).
	
	\bibitem{seiler16a}
	{\mbox{Ch}}. Seiler, J.A. Agner, P. Pillet and F. Merkt,  J. Phys. B: At. Mol.
	Opt. Phys.  \textbf{49}, 094006 (2016).
	
	\bibitem{motsch14a}
	M. Motsch, P. Jansen, J.A. Agner, H. Schmutz and F. Merkt,  Phys. Rev. A
	\textbf{89}, 043420 (2014).
	
	\bibitem{meek08a}
	S.A. Meek, H.L. Bethlem, H. Conrad and G. Meijer,  Phys. Rev. Lett.
	\textbf{100}, 153003 (2008).
	
	\bibitem{meek09a}
	S.A. Meek, G. Santambrogio, H. Conrad and G. Meijer,  {J. Phys. Conf. Ser.}
	\textbf{194}, 012063 (2009).
	
	\bibitem{hogan13a}
	S.D. Hogan, {\mbox{Ch}}. Seiler and F. Merkt,  J. Phys. B: At. Mol. Opt. Phys.
	\textbf{46}, 045303 (2013).
	
	\bibitem{seiler11a}
	{\mbox{Ch}}. Seiler, S.D. Hogan, H. Schmutz, J.A. Agner and F. Merkt,  Phys.
	Rev. Lett.  \textbf{106}, 073003 (2011).
	
	\bibitem{zimmerman79a}
	M.L. Zimmerman, M.G. Littman, M.M. Kash and D. Kleppner,  Phys. Rev. A
	\textbf{20}, 2251--2275 (1979).
	
	\bibitem{drake91a}
	G.W.F. Drake and R.A. Swainson,  {Phys. Rev. A}  \textbf{44}, 5448--5459 (1991).
	
\end{thebibliography}

\end{document}